\documentstyle[12pt]{article}
\def\1{{\sf{\setbox0=\hbox{1}1\kern-.7\wd0\hbox{1}}}}
\def\thebibliography#1{
{\large \bf References}%
\list{\arabic{enumi})}%
  {\settowidth{\labelwidth}{#1)}%
   \leftmargin=\labelwidth
   \advance \leftmargin by \labelsep
   \usecounter{enumi}}%
\def\newblock{\hskip .11em plus .33em minus .07em}%

\sloppy
\clubpenalty=4000 \widowpenalty=4000
 \sfcode`\.=1000\relax}%

\def\R{{\rm{\setbox0=\hbox{R}R\kern-1.2\wd0\hbox{I}
\vbox{\hrule height0ex width.7\wd0}}}}

\def\C{{\rm{\setbox0=\hbox{C}C\kern-.65\wd0\vbox{\hrule
height1.37ex width.06\wd0\kern.09ex\hrule height0ex width.7\wd0}}}}

\def\Z{{\sf{\setbox0=\hbox{Z}Z\kern-.7\wd0\hbox{Z}}}}

\def\<{\langle}
\def\>{\rangle}
\catcode`@=11
\newskip\ccentering \ccentering=0pt plus 1000pt minus 1000pt
\def\openup{\afterassignment\@penup\dimen@=}
\def\@penup{\advance\lineskip\dimen@
  \advance\baselineskip\dimen@
  \advance\lineskiplimit\dimen@}
\def\eqalign#1{\null\,\vcenter{\openup\jot\m@th
  \ialign{\strut\hfil$\displaystyle{##}$&$\displaystyle{{}##}$\hfil
      \crcr#1\crcr}}\,}
\newif\ifdt@p
\def\displ@y{\global\dt@ptrue\openup\jot\m@th
  \everycr{\noalign{\ifdt@p \global\dt@pfalse
      \vskip-\lineskiplimit \vskip\normallineskiplimit
      \else \penalty\interdisplaylinepenalty \fi}}}
\def\@lign{\tabskip\z@skip\everycr{}} % restore inside \displ@y
\def\displaylines#1{\displ@y
  \halign{\hbox to\displaywidth{$\@lign\hfil\displaystyle##\hfil$}\crcr
    #1\crcr}}
\def\eqalignno#1{\displ@y \tabskip\ccentering
  \halign to\displaywidth{\hfil$\@lign\displaystyle{##}$\tabskip\z@skip
    &$\@lign\displaystyle{{}##}$\hfil\tabskip\ccentering
    &\llap{$\@lign##$}\tabskip\z@skip\crcr
    #1\crcr}}
\def\leqalignno#1{\displ@y \tabskip\ccentering
  \halign to\displaywidth{\hfil$\@lign\displaystyle{##}$\tabskip\z@skip
    &$\@lign\displaystyle{{}##}$\hfil\tabskip\ccentering
    &\kern-\displaywidth\rlap{$\@lign##$}\tabskip\displaywidth\crcr
    #1\crcr}}

\def\otop{%
	\mbox{\footnotesize $ \,%
			\bigcirc \kern-0.91em \lower0.18em\hbox {$\top$}	\: $}%
		  }
\def\obot{%
	\mbox{\footnotesize $ \,%
			\bigcirc \kern-0.91em \hbox {$\bot$}	\: $}%
		  }
\def\[[{ [\! [}
\def\]]{ ]\! ]}

\begin{document}
\thispagestyle{empty}
\setcounter{page}{0}
\baselineskip=20pt
\begin{flushright}
 TOYAMA-80 \\
 KANAZAWA-94-21 \\
 October, 1994
\end{flushright}
\vspace*{0.5cm}
\begin{center}
{\Large \bf Distance Formula for Grassmann Manifold \\}
{\large---Applied to Anandan--Aharonov Type Uncertainty Relation---}
\vspace{2.5cm}\\
Minoru HIRAYAMA        \footnote{e-mail address: hirayama@jpntyavm.bitnet}
, Takeshi HAMADA$^*$   \footnote{e-mail address: hamada@hep.s.kanazawa-u.ac.jp}
 and Jin CHEN
\vspace{1.3cm}\\
Department of Physics\\
Toyama University\\
Toyama 930, Japan
\vspace{1cm}\\
${}^*$Department of Physics\\
Kanazawa University\\
Kakuma, Kanazawa 920-11, Japan
\end{center}
\vspace{2.5cm}
\noindent
{\large Abstract}
\vspace{0.5cm}

The time-energy uncertainty relation of Anandan-Aharonov is
generalized to a relation involving a set of quantum state vectors.
This is achieved by obtaining an explicit formula for the
distance between two finitely separated points in the
Grassmann manifold.
%
%   1-1
%
\newpage
\noindent
{\bf $\S$1. Introduction}
\vspace{0.5cm}

We begin with briefly reviewing the conventional time-energy uncertainty
relation in quantum mechanics. Let $A$ be an ovservable without
explicit time-dependence and $|\psi(t)\>$ be a normalized
quantum state vector obeying the Schr\"odinger equation
with a hermitian Hamiltonian $H$. If we define $\Delta A$ and
$\tau_A$ by
$$
\eqalignno{
 &\Delta A= \sqrt{ \< \psi(t)|A^2|\psi(t) \> - \< \psi(t)|A|\psi(t) \> ^2}\ ,
  &(1.1)\cr
 &\tau_A=\left| \frac d{dt} \< \psi(t)|A|\psi(t) \> \right|^{-1} \Delta A \ ,
  &(1.2)\cr}$$
and take the equation
$$
\frac d{dt} \< \psi(t)|A|\psi(t) \> =\frac 1{i\hbar} \< \psi(t)|\ [A,H]\
|\psi(t) \>
\eqno(1.3)$$
into account, we are led to the uncertainty relation$^{1)}$
$$
\tau_A \Delta H \geq \frac {\hbar}2 \ . \eqno(1.4)$$
The quantity $\tau_A$ is interpreted as the time necessary
%
% 1-2
%
for the distribution of $ \< \psi(t)|A|\psi(t) \> $ to be recognized
to have clearly changed its shape.

In contrast with the well known result given above,
Anandan and Aharonov have recently succeeded in obtaining
quite an interesting inequality.$^{2)}$ They consider the
case that the $|\psi(t) \> $ develops in time obeying
$$
\eqalignno{
& i\hbar \frac d{dt}|\psi(t) \> =H(t)|\psi(t) \> \ , &(1.5) \cr
&  \< \psi(t)|\psi(t) \> =1 \ , &(1.6)\cr }$$
where $H(t)$ is an operator which is hermitian and might
be time-dependent. They conclude that
$$
\int^{t_2}_{t_1} \Delta {\cal E}(t)dt \geq \hbar \mbox{Arccos}
(| \< \psi(t_1)|\psi(t_2) \> |)\ ,\eqno(1.7)$$
where $\Delta {\cal E}(t)$ is given by
$$
\Delta {\cal E}(t)=\sqrt{ \< \psi(t)|H(t)^2|\psi(t) \> -
 \< \psi(t)|H(t)|\psi(t) \> ^2} \ .\eqno(1.8)$$
The inequality (1.7), which we refer to as the
%
% 1-3
%
Anandan-Aharonov time-energy uncertainty relation, has been
derived through a geometrical investigation of the set of
normalized quantum state vectors. The r.h.s.\ of (1.7) can
be regarded as the distance between two points
in a complex projective space.$^{3)}$
We note that Montgomery also proposed an interesting time-energy
uncertainty relation of a similar nature.$^{4)}$

In this paper, we seek the generalized version of (1.7).
We consider a set of $N$ orthonormal vectors
$ \{|\psi_i(t) \> :i=1,2,\ldots,N \}$ satisfying
$$
 \< \psi_i(t)|\psi_j(t) \> =\delta_{ij}\ ,\quad i,j=1,2,\ldots,N, \eqno(1.9)$$
each of which obeying the Schr\"odinger equation (1.5).
We define $N\times N$ matrices $A(t_1,t_2)$ and $K(t_1,t_2)$ by
$$
\eqalignno{
& A(t_1,t_2)=\left(a_{ij}(t_1,t_2)\right)\ ,\quad
  a_{ij}(t_1,t_2)= \< \psi_i(t_1)|\psi_j(t_2) \> \ , &(1.10) \cr
& K(t_1,t_2)=A^\dagger (t_1,t_2)A(t_1,t_2) &(1.11) \cr}$$
and $\kappa_i(t_1,t_2),i=1,2,\ldots,N,$ to be the eigenvalues
%
% 1-4
%
of $K(t_1,t_2)$. Defining the generalization of (1.8) by
$$
\Delta {\cal E}_N(t)=\sqrt{ \sum_{i=1}^N  \< \psi_i(t)|H(t)^2|\psi_i(t) \> -
\sum_{i,j=1}^N | \< \psi_i(t)|H(t)|\psi_j(t) \> |^2}\ ,\eqno(1.12)$$
we find that $\Delta {\cal E}_N(t)$ satisfies
$$
\int_{t_1}^{t_2}\Delta {\cal E}_N(t)dt \geq \hbar \sqrt{ \sum_{i=1}^N
\left\{ \mbox{Arccos}\sqrt{\kappa_i(t_1,t_2)} \right\}^2}\ . \eqno(1.13)$$
The inequality (1.13) can be written in an operator form as
$$
\eqalign{
\int_{t_1}^{t_2} &\sqrt{\mbox{Tr}( P(t)[ H(t),[ H(t),P(t)]])} \mbox{dt} \cr
&\geq \sqrt2 \hbar
\sqrt{ \mbox{Tr}(\{\mbox{Arccos} \sqrt{P(t_1)P(t_2)} \}^2 )}\  ,} \eqno(1.14)$$
where $P(t)$ is defined by
$$
P(t)=\sum_{i=1}^N|\psi_i(t) \>  \< \psi_i(t)|,  \eqno(1.15)$$
and Tr denotes the trace in the Hilbert space. The result (1.13) is
obtained through a geometrical investigation
of the Grassmann manifold $G_N$ mentioned below.
%%
%%  The inequality (1.14) can be rewritten as
%%  $$
%%  \int_{t_1}^{t_2} \left\| \frac{dP(t)}{dt} \right\| dt \geq
%%  \sqrt{2} \left\| \mbox{Arccos} \sqrt{P(t_1)P(t_2)} \right\|,
%%  \eqno(1.16)
%%  $$
%%  where $\| A \|$ denotes the Hilbert-Schmidt norm of an operator A.

This paper is organized as follows. In $\S2$, we introduce
some objects such as geodesic, distance, etc., defined on
the set, $G_N$, of $N$--dimensional linear subspaces of a
Hilbert space. In $\S3$, we obtain an explicit formula for
the distance between two points in $G_N$.
In $\S4$, we discuss that the distance introduced above satisfies
the properties of distance including the triangle inequality.
The inequality (1.13) is derived in $\S5$. The final
section, $\S6$, is devoted to discussions. Some appendices
are attached to explain some necessitated calculations.
\newpage
\noindent
{\bf $\S2$. General discussions of the distance in $G_N$}
\vspace{0.3cm}\\
{\bf 2.1 \ $N$-th Grassmannian}
\vspace{0.3cm}

Given a Hilbert space h,we consider vectors
$|\psi_i \> ,i=1,2,\ldots,N$,belonging to h and satisfying
$ \< \psi_i|\psi_j \> =\delta_{ij}$. We call the set
$$\Psi=(|\psi_1 \> ,|\psi_2 \> ,\ldots,|\psi_N \> )\eqno(2.1)$$
an $N$-frame of h and the set
$$[\Psi] =\{\Psi u: u \in U(N)\} \eqno(2.2)$$
an $N$-plane of h,where $\Psi u$ is defined by
$$\Psi u=(\sum_{i=1}^N |\psi_i \> u_{i1},
\sum_{j=1}^N |\psi_j \> u_{j2},\ldots,
\sum_{k=1}^N |\psi_k \> u_{kN}).\eqno(2.3) $$
It is clear that the $[\Psi]$ and the projection operator
$$ P = \sum_{i=1}^N |\psi_i \>  \< \psi_i| \eqno(2.4)$$
are invariant under the replacement $\Psi \to \Psi u$.
We denote the set of all the $\Psi$'s of h by $S_N$.
Then the set $G_N$ defined by
$$G_N=\{ [\Psi]:\Psi \in S_N\} \eqno(2.5)$$
is known to constitute a manifold of complex dimension
$N(\mbox{dim h}-N)$.
We hereafter call $G_N$ the $N$-th Grassmann manifold,
or simply the $N$-th Grassmannian.
\vspace{1cm}\\
\noindent
{\bf 2.2 \  Geodesics in the set of unitary operators on h}
\vspace{0.5cm}

We denote the set of unitary operators on h by $\Omega$.
A local coordinate of $\Omega$ is denoted by
$s=(s^1, s^2, s^3, \ldots), s^1, s^2, s^3, \ldots \in \R$.
We define the infinitesimal distance between two unitary operators
$W$ and $W+dW$ by
$$D(W, W+dW)=\| dW \|,\eqno(2.6)$$
where the Hilbert--Schmidt norm
$\| w\|$ of an operator $w$ is defined by
$$\| w\| = \sqrt{{\rm Tr}(w w^\dagger )},\eqno(2.7)$$
{\rm Tr} denoting the trace on h.
If we define $g_{\mu\nu}(s)$ by
$$ g_{\mu\nu}(s)=\mbox{Re}\{{\rm Tr} (\partial_\mu W \partial_\nu W^\dagger)\}
,\partial_\mu = \frac{\partial}{\partial s^\mu},\eqno(2.8)$$
we have
$$D(W, W+dW)=\sqrt{g_{\mu\nu}(s)ds^\mu ds^\nu}.\eqno(2.9)$$
The $g_{\mu\nu}(s)$ defined above can be regarded as
the metric tensor of the set $\Omega$ of unitary operators on h.
It is evident that $g_{\mu\nu}(s)$ transforms as a tensor under the
transformation of the coordinate $s$
and the value of the
infinitesimal destance $D(W, W+dW)$ is independent of the
choice of $s$.
Given the metric
$g_{\mu\nu}(s)$, a geodesic of $\Omega$ is
defined as a solution
$s(t)=(s^1(t),s^2(t),\ldots), t \in \R$, of the equation
$$\ddot s^\mu(t)+ \Gamma^\mu_{\lambda\sigma}(s(t))\dot
s^\lambda (t) \dot s^\sigma (t) = 0, \quad \mu =1,2,\ldots,
\eqno(2.10)$$
$$\Gamma^\mu_{\lambda\sigma}(s) =\frac12 g^{\mu\nu}(s)
\{\partial_\lambda g_{\sigma\nu}(s) +
\partial_\sigma g_{\lambda\nu}(s) -
\partial_\nu g_{\lambda\sigma}(s)\},\eqno
(2.11)$$
where ($g^{\mu\nu}(s)$) is the inverse of ($g_{\mu\nu}(s)$)
and the dot denotes the derivative with respect to t.

The one parameter subgroup $\{e^{itY}:Y=Y^\dagger,t\in \R \}$
should be more or less related to
a geodesic of $\Omega$ since we have
$$\eqalign{
\frac{d}{dt}\left\{\left\|\frac{dW(t)}{dt}\right\|^2\right\}
&=
\frac{d}{dt}\{g_{\mu\nu}(s(t))\dot s^\mu(t)\dot s^\nu(t)\}\cr
&=
2g_{\mu\nu}(s(t))\dot s^\nu(t)\{\ddot s^\mu(t)
+ \Gamma^\mu_{\lambda\sigma}(s(t))\dot s^\lambda(t)
\dot s^\sigma(t) \} }\eqno(2.12)
$$
and
$$\frac{d}{dt}\left\{\left\|\frac{d}{dt}e^{itY} \right\|^2\right\}
 = \frac{d}{dt}\{\|Y\|^2\}=0. \eqno(2.13)$$
The fact is that any geodesic in
$\Omega$ passing the point $\1$ (unit operator)
can be regarded as a one parameter subgroup of the above form.
Although this fact can be seen in mathematical literatures,${}^{5)}$
we discuss it in the Appendix A for self-containedness.
We note that the length of the geodesic
$\{e^{itY}:0 \leq t \leq 1, Y=Y^\dagger \}$ connecting the two points
$\1$ and $e^{iY}$
in $\Omega$ is given by
$$\int_0^1 \left\|\frac{d}{dt}e^{itY}\right\|dt = \|Y\|. \eqno(2.14)$$
\newpage
\noindent
{\bf 2.3 \  Distance in $G_N$}
\vspace{0.3cm}

To an $N$-frame $\Psi(t) = (|\psi_1(t) \> ,|\psi_2(t) \> ,\ldots,|\psi_N(t) \>
)
\in S_N,0\leq t\leq 1$,
there correspond an $N$-plane $[\Psi(t)]\in G_N$ and a projection operator
$P(t)= \sum_{i=1}^N|\psi_i(t) \>  \< \psi_i(t)|$.
Since the eigenvalues of $P(1)$ are equal to those of $P(0)$
including multiplicities,there exist a unitary operator $W$ such that
$$P(1) = W^\dagger P(0) W, \quad W=e^{iY}, \quad Y^\dagger = Y. \eqno(2.15)$$
The discussion of the previous subsection suggests that
we might be able to define the distance $d([\Psi(0)],[\Psi(1)])$
between two points $[\Psi(0)]$ and $[\Psi(1)]$ of the Grassmannian $G_N$
by the formula similar to the r.h.s.\ of (2.14). We define it by
$$d([\Psi(0)],[\Psi(1)])=\mathop{\mbox{Min}}_{Y\in \Sigma} \|Y\|\eqno(2.16)$$
where $\Sigma$ is the set of hermitian operators specified by
$P(0)$ and $P(1)$ in the following way:
$$ \Sigma=\{ Y:Y=Y(P(0),P(1))=-Y(P(1),P(0))=Y^\dagger,
e^{-iY}P(0)e^{iY}=P(1)\}.\eqno(2.17)$$
Supposing that Min $\| Y \|$, $Y \in \Sigma$, is attained by
$Y_0 \in \Sigma$, we have
$$d([\Psi(0)],[\Psi(1)])=\| Y_0 \|. \eqno(2.18)$$
As will be shown later,
if we require that the functional form of
$Y_0(P(0), P(1))$ should be fixed independently of the choice of
$P(0)$ and $P(1)$, the $Y_0(P(0),P(1))$
is determined uniquely.
It should be stressed that the r.h.s.\ of (2.16) or (2.18)
is invariant under the replacement $\Psi\rightarrow\Psi u,
u \in U(N)$ and can be regarded as a quantity
defined on $G_N$.
After obtaining the explicit expressions of $Y_0$ and
$d([\Psi(0)],[\Psi(1)])$ in \S3, we will discuss in $\S4$
that the above defined distance in $G_N$ satisfies
the property of distance:
$$d([\Psi],[\Phi])=d([\Phi],[\Psi]) \geq 0,\eqno(2.19)$$
$$d([\Psi],[\Phi])=0 \iff [\Psi]=[\Phi], \eqno(2.20)$$
$$d([\Psi],[\Phi])\leq d([\Psi],[\Xi])+d([\Xi],[\Phi]),
\eqno(2.21)$$
for any $[\Psi],[\Phi],[\Xi] \in G_N$.
%
%
%
% 3-1
%
\newpage
\noindent
{\bf $\S$3. Explicit formula for the distance in $G_N$}
\vspace{0.5cm}

Adopting abbreviated notations $P_0$ and $P_1$ for the projection
operators $P(0)$ and $P(1)$, respectively, we have
$$
P_m^2=P_m,\quad (P_m)^\dagger=P_m,\quad m=0,1. \eqno(3.1)$$
It is not difficult to see that the most general form of
$Y$ satisfying $Y=Y(P_0,P_1)=-Y(P_1,P_0)=Y^\dagger$ is given by
$$
Y=\alpha(P_0 P_1)P_0-\alpha(P_1 P_0)P_1+i
\{\beta(P_0 P_1)P_0 P_1-\beta(P_1 P_0)P_1 P_0 \}, \eqno(3.2)$$
where $\alpha(z)$ and $\beta(z)$ are real analytic functions of
$z$ involving no inverse powers of $z$. As is explained
in the Appendix B, we obtain
$$
\mbox{Tr}(e^{iY} P_1 e^{-iY} P_0)=\sum_{i=1}^N
\kappa_i \left\{ \cos\sqrt{s(\kappa_i)}-(1-\kappa_i)\beta(\kappa_i)
\frac{\sin\sqrt{s(\kappa_i)}}{\sqrt{s(\kappa_i)}} \right\}^2,\eqno(3.3)$$
where $\kappa_i,i=1,2,\ldots,N,$ are the eigenvalues of the
$N \times N$ matrix $K$ whose $ij$--element is given by
$$
K_{ij}=\sum_{k=1}^N \< \psi_i(0)|\psi_k(1) \>  \< \psi_k(1)|\psi_j(0) \>
.\eqno(3.4)$$
%
% 3-2
%
The $s(\kappa)$ in (3.3) is calculated from (B.23) and (B.25) to be
$$
s(\kappa)=(1-\kappa)(\{\alpha(\kappa)\}^2 + \kappa \{\beta(\kappa)\}^2).
\eqno(3.5)$$
In the Appendix C, the eigenvalue $\kappa_i$ is shown to satisfy
$$
0\leq \kappa_i \leq 1,\quad i=1,2,\ldots,N. \eqno(3.6)$$
Another requirement $e^{-iY}P_0e^{iY}=P_1$ characterizing
$Y \in \Sigma$ is equivalent to
$$
\mbox{Tr}(e^{iY}P_1e^{-iY} P_0)=N \eqno(3.7)$$
as is seen from
$$
\|e^{-iY}P_0 e^{iY}-P_1\|^2=2\{ N-\mbox{Tr}(e^{iY}P_1 e^{-iY} P_0)\}.
\eqno(3.8)$$
{}From (3.3) and (3.7), we obtain the condition to
specify the functions $\alpha(z)$ and $\beta(z)$ defining $Y\in \Sigma$:
$$
\sum_{i=1}^N C(\kappa_i)\cos^2 (\sqrt{s(\kappa_i)}-\phi(\kappa_i))=N,
\eqno(3.9)$$
where $C(\kappa)$ and $\phi(\kappa)$ are given by
$$
C(\kappa)=\kappa \frac{\{\alpha(\kappa)\}^2+\{\beta(\kappa)\}^2}
{\{\alpha(\kappa)\}^2+\kappa \{\beta(\kappa)\}^2}\ , \eqno(3.10)$$
$$
\tan \phi(\kappa)=-\frac{\sqrt{1-\kappa}\beta(\kappa)}
{\sqrt{\{\alpha(\kappa)\}^2+\kappa\{\beta(\kappa)\}^2}}\ . \eqno(3.11)$$
%
% 3-3
%
Except for the trivial case $\kappa_1=\kappa_2=\cdots=\kappa_N=1$,
the condition (3.9) is realized only when $\alpha(\kappa_i)=0,
i=1,2,\ldots,N$.  Since we are considering generic cases
where $\kappa_i$'$s$ satisfy (3.6)
and requiring that the functional form of $\alpha(z)$
is independent of $P_0$ and $P_1$, we conclude
that $\alpha(z)$ vanishes identically :
$$
\alpha(z)=0. \eqno(3.12)$$
Then $C(\kappa)$ equals 1 and we see from (3.5), (3.9) and
(3.11) that $\beta(\kappa)$ should satisfy
$$
\tan\phi(\kappa)=-\sqrt{\frac{1-\kappa}{\kappa}}\mbox{sgn}
(\beta(\kappa)) \eqno(3.13)$$
and
$$
\frac1{\pi} \{\phi(\kappa)-\sqrt{\kappa(1-\kappa)}|\beta(\kappa)| \}
\in \Z. \eqno(3.14)$$
Eliminating $\phi(\kappa)$ from (3.13) and (3.14), we are
led to the condition
$$
\tan(\sqrt{z(1-z)}\beta(z))=-\sqrt{\frac{1-z}z} \eqno(3.15)$$
to determine $\beta(z)$.  We have seen that the operator $Y$
belonging to $\Sigma$ of (2.17) is specified by (3.2), (3.12)
and (3.15).

We next determine $Y_0$ which attains the minimum of
$\| Y \|, Y \in \Sigma$. It is
%
% 3-4
%
now easy to observe that $Y_0$ is uniquely given by
$$
\eqalign{
Y_0 &=i \{\beta_0(P_0 P_1)P_0 P_1-\beta_0(P_1 P_0)P_1 P_0 \}, \cr
&\equiv G(P_0, P_1)} \eqno(3.16)$$
where $\beta_0(z)$ is given by
$$
\eqalign{
\beta_0(z)&=-\frac1{\sqrt{z(1-z)}} \mbox{Arctan}
            \sqrt{\frac{1-z}{z}} \cr
 &=-\frac1{\sqrt{z(1-z)}} \mbox{Arccos} \sqrt{z} \cr
 &=-\sum_{n=0}^\infty \frac{(2n)!!}{(2n+1)!!}(1-z)^n, \quad
 |1-z| < 1.}\eqno(3.17)$$
With the help the relation $(P_0-P_1)^{2n}P_0P_1=(1-P_0P_1)^n
P_0P_1, n=0,1,2,\ldots,$ we have
$$
G(P_0, P_1)=\frac 1i \sum_{n=0}^{\infty}
\frac{(2n)!!}{(2n+1)!!}(P_0-P_1)^{2n}(P_0P_1-P_1P_0). \eqno(3.18)$$
The distance $d([\Psi(0)],[\Psi(1)])$ is calculated by
(2.16).  Through procedures similar to those of
Appendix B, we obtain
$$
\eqalign{
d([\Psi(0)],[\Psi(1)])&=\sqrt{2\mbox{Tr}((x-x^2)\{\beta_0(x)\}^2)},
 \quad x=P_0 P_1, \cr
&=\sqrt{2\sum_{i=1}^N(\mbox{Arccos} \sqrt{\kappa_i})^2}. }\eqno(3.19)$$
In the case of $N=1$, the above distance reduces
to $\sqrt2 \mbox{Arccos}| \< \psi(0)|\psi(1) \> |$ and reproduces the
distance of the complex projective space,$^{3)}$ which was utilized by
Anandan and Aharonov.$^{2)}$
After some manipulations, we obtain an important relation
$$
\|[G(P_0, P_1), P_0]\|=\|G(P_0, P_1)\|, \eqno(3.20)$$
which will be useful for later discussions.

In a recent paper, Avron, Seiler and Simon$^{6)}$ argued algebraic
properties of a pair of projection operators. They noted that,
under a suitable condition on the norm of
$P_0-P_1$, the unitary operator
$$
U=\frac{P_0P_1+(1-P_0)(1-P_1)}{\sqrt{1-(P_0-P_1)^2}} \eqno(3.21)$$
transforms $P_0$ into $P_1$ :
$$
U^\dagger P_0 U=P_1. \eqno(3.22)$$
The expression on the r.h.s. of (3.21) makes sense since
$(P_0-P_1)^2$ commutes with $P_0$ and $P_1$. It can be read off
from Ref. 6) that (3.21) was originally obtained by T. Kato
many years ago. In the Appendix D, it is shown that our
$\exp(iG(P_0, P_1))$ is nothing but the $U$ of (3.21) :
$$
e^{iG(P_0, P_1)}=U. \eqno(3.23)$$
%
% 4-1
%
\newpage
\noindent
{\bf $\S$.4 Property of distance}
\vspace{0.5cm}

We here discuss how the properties (2.19), (2.20) and (2.21)
are assured. It is evident that the $d([\Psi], [\Phi])$ defined by
(2.16) staisfies (2.19). Noticing that, in the case that
$[\Psi]=[\Phi]$, all the eigenvalues corresponponding to $\kappa_i$'s
of $\S 3$ are equal to 1 and all the $\mbox{Arccos}\sqrt{\kappa_i},
i=1,2,\ldots,N,$ vanish, we see that (2.20) is also satisfied.

We now go on to the triangle inequality (2.21). For the case of
$N=1$, (2.21) is equivalent to
$$ \mbox{Arccos}\sqrt\kappa \leq \mbox{Arccos}\sqrt\sigma +
\mbox{Arccos}\sqrt\lambda, \eqno(4.1)$$
where $\kappa, \lambda$ and $\sigma$ are defined by
$$
\kappa=| \< \phi|\psi \> |^2,\quad \sigma=| \< \psi|\xi \> |^2,\quad
\lambda=| \< \xi|\phi \> |^2 \eqno(4.2)$$
with $|\phi \> , |\psi \> $ and $|\xi \> $ being unit vectors.
The inequality (4.1) can be proved algebraically as is seen
in Appendix E. In general cases of $N \geq 2$, we have not
succeeded in proving (2.21) algebraically
%
% 4-2
%
relying solely upon the formula (3.19). In the following,
we describe an analytic proof of (2.21).

We consider three $N$-planes $[\Psi], [\Xi], [\Phi] \in G_N$
and denote the projection operators corresponding to them by
$P_\Psi, P_\Xi$ and $P_\Phi$, respectively
$(P_\Psi=\sum_{i=1}^N|\psi_i \>  \< \psi_i|, etc.)$.
We define hermitian operators $Y_1, Y_2$ and $Y_3$ by
$$
Y_1=-G(P_\Psi, P_\Phi), \quad Y_2=-G(P_\Psi, P_\Xi),
\quad Y_3=-G(P_\Xi, P_\Phi), \eqno(4.3)$$
where $G(P_0, P_1)$ is defined by (3.16). A piecewise
smooth geodesic $\Gamma=\{ \gamma(t) : 0\leq t \leq 1\}$
in $\Omega$ is defined as
$$ \gamma(t)= \left \{
\eqalign{
 &\exp \left(i \frac{t}{t_0}Y_2 \right); \quad 0 \leq t \leq t_0 \cr
 &\exp \left(i \frac{t-t_0}{1-t_0}Y_3 \right)e^{iY_2};
 \quad t_0 \leq t \leq 1, } \right. \eqno(4.4) $$
where $t_0$ is a constant satisfying $0 < t_0 < 1$.

We define a set of projection operators by
$$
\wedge_N = \{P: P^2=P, P^\dagger=P, \mbox{Tr}P=N \} \eqno(4.5)$$
and consider a path
$$
C=\{P(t): 0\leq t \leq 1 \} \subset \wedge_N, \eqno(4.6)$$
where $P(t)$ is given by
$$
P(t) = \gamma(t) P_\Psi \gamma(t)^\dagger. \eqno(4.7)$$
The projection operator $P(t)$ satisfies $P(0)=P_\Psi,
P(t_0)=P_\Xi$ and $P(1)=P_\Phi$. The length of $C$ is defined by
$$
l(C)=\int_C \|dP\|= \int_0^1 \left\| \frac{dP(t)}{dt}
\right\| dt. \eqno(4.8)$$
Then we have
$$
\eqalign {
l(C) &=\int_0^{t_0}\frac1{t_0}\|[Y_2, P_\Psi]\| dt+ \int_{t_0}^1
\frac1{1-t_0}\|[Y_3, P_\Xi]\| dt \cr
&=\|[Y_2, P_\Psi]\| + \|[Y_3, P_\Xi]\| \cr
&=\|Y_2\| + \|Y_3\| \cr
&=d([\Psi], [\Xi]) + d([\Xi], [\Phi]),} \eqno(4.9)$$
where the relation (3.20) has been made use of. Similarly, the length
of the path
$$
C_0=\{P_0(t) : 0 \leq t \leq 1\} \subset \wedge_N, \eqno(4.10)$$
with
$$
P_0(t)= e^{itY_1} P_\Psi e^{-itY_1} \eqno(4.11)$$
is given by
$$
l(C_0)=\int_{C_0}\|dP\|= \int_0^1 \left\| \frac{dP_0(t)}{dt}\right\|
dt=\|[Y_1, P_\Psi]\|=\|Y_1\|=d([\Psi], [\Phi]). \eqno(4.12)$$
The metric tensor $G_{\mu\nu}(s)$ corresponding to an
infinitesimal distance $\|dP\|$ is given by
$$
\|dP\| = \sqrt{G_{\mu\nu}(s)ds^\mu ds^\nu}, \eqno(4.13)$$
$$
G_{\mu\nu}(s)=\mbox{Tr}(\partial_\mu P(s)\partial_\nu P(s)),\quad
\partial_\mu=\frac{\partial}{\partial s^\mu}, \eqno(4.14)$$
where $s$ denotes a set of local coordinates of $\wedge_N$.
The $l(C_0)$ is stationary under any variation $\delta$ of
$C_0$ with the end points $P_\Psi$ and $P_\Phi$ fixed.
This is because the length $l(C_0)$ is a function only of
$P_\Psi$ and $P_\Phi$. On the other hand, for some choices of
$P_\Xi(e.g., \< \xi_i | \psi_j \> = \< \xi_i | \phi_j \> =0,
i,j=1,2,\ldots,N)$,
the $l(C)$ can be clearly larger than $l(C_0)$.
For generic $P_\Xi$, the $l(C)$ is not stationary under the above
mentioned variations $\delta$ since the value of $l(C)$ varies
continuously with $P_\Xi$.
For some choices of $P_\Xi(e.g., P_\Xi=P_\Phi \ \mbox{or} \ P_\Xi=P_\Psi)$,
the $l(C)$ coincides with $l(C_0)$. From these discussions, we see
$l(C) \geq l(C_0)$, implying (2.21).

Note that the $Q(t)$ defined by
$$
Q(t)=tP_\Phi + (1-t)P_\Psi, \quad 0\leq t \leq 1, \eqno(4.15)$$
satisfies $\frac d{dt} \left\| \frac{dQ(t)}{dt}
\right\|^2=0$ and
$\delta(l(Q))=0$  with $l(Q)$ given by $l(Q)\equiv \int_0^1
\left\| \frac{dQ(t)}{dt} \right\| dt
=\| P_\Phi-P_\Psi \| \leq l(C_0)$.
The set $\{ Q(t): 0\leq t \leq 1\}$, however, does not belong to
$\wedge_N$ since $\{Q(t)\}^2$ is different from $Q(t)$ for
general values of $t$. So, we should not regard
$\|P_\Phi-P_\Psi \|$ as the distance in $\wedge_N$ between
$P_\Psi$ and $P_\Phi$ although we have $\| P_\Phi-P_\Psi \| \leq
\| P_\Phi-P_\Xi \|+\|P_\Xi-P_\Psi \|$.

Finally, we mention one more inequality. If we define a hermitian
operator $Y(t)$ by
$$
\gamma(t)=e^{iR(t)Y(t)}, \quad 0\leq t \leq 1, \eqno(4.16)$$
$$
R(t) \in \R, \quad R(0)=0, \quad R(1)=1, \eqno(4.17)$$
$$
\| Y(t) \|= \mbox{const.} =\| Y(1) \|, \eqno(4.18)$$
the relation $e^{iY_3} e^{iY_2}=\gamma(1)=e^{iY(1)}$ and the
Campbell--Hausdorff formula yield
$$
iY(1)=iY_3+iY_2+\frac{i^2}2 [Y_3, Y_2]+\frac{i^3}{12}
([[Y_3, Y_2], Y_2]-[[Y_3, Y_2], Y_3]) + \cdots. \eqno(4.19)$$
The path $C_1 \equiv \{P_1(t): P_1(t)=e^{itY(1)}P_\Psi e^{-itY(1)} \}$
also connects $P_\Psi$ with $P_\Phi$ in $\wedge_N$.
The $P_1(t)$ satisfies $\frac d{dt}
\left\| \frac{dP_1(t)}{dt}\right\|^2=0 $
which is necessary but not sufficient for $C_1$ to be a geodesic.
The $l(C_1)$ for generic $P_\Xi$, however, would not be
stationary under the variations $\delta$ mentioned above.
Discussions similar to those of the case of $l(C)$ lead us to
$\| Y(1)\|=l(C_1) \geq l(C_0)=\|Y_1\|$. On the other hand, as is
seen in the Appendix F, we have
$$
\left\| \frac{d\gamma(t)}{dt}\right\| \geq \left|
\frac{dR(t)}{dt} \right| \|Y(1)\|. \eqno(4.20)$$
{}From (4.8), (4.9), $\left\| \frac{dP(t)}{dt} \right\|
=\left\| \frac{d\gamma(t)}{dt}\right\|$ implied by (3.20), (4.17)
and $\|Y(1)\| \geq \| Y_1 \|$, we obtain
$$
\|Y_2\|+\|Y_3\| \geq \|Y(1)\| \geq \|Y_1\|, \eqno(4.21)$$
which somewhat reveals more detailed feature of
$\|Y_2\|+\|Y_3\| \geq \|Y_1\|$.
%
% 5-1
%
\newpage
\noindent
{\bf $\S$5. Time-energy uncertainty relation}
\vspace{0.5cm}

Having obtained the explicit formula (3.19) for the distance in
$G_N$, we here derive the uncertainty relation (1.13)
or (1.14).  The projection operator $P(t)$ is defined by
(1.15) and $|\psi_i(t) \> ,i=1,2,\ldots,N,$ develops in
time obeying (1.5).  We then have
$$\eqalign{
P(t+dt)= P(t)+&\frac{dt}{i\hbar}[H(t),P(t)] \cr
&+\frac{(dt)^2}{2(i\hbar)^2} \left\{ i\hbar[\frac{dH(t)}{dt},P(t)]+
[H(t),[H(t),P(t)]] \right\} + \cdots. }\eqno(5.1)$$
When $[\Psi(0)]$ and $[\Psi(1)]$ in (3.19) are close to
each other, $\kappa_i, i=1,2,\ldots,N,$ are nearly equal to 1.
Noticing that $(\mbox{Arccos} \sqrt{\kappa})^2 \approx 1-\kappa$
for $\kappa \approx 1$, we see
$$
d([\Psi(t)],[\Psi(t+dt)])\approx \sqrt{2\sum_{i=1}^N(1-\kappa_i(t)),}
\eqno(5.2)$$
where $\kappa_i(t)$'$s$ are obtained from $P(t)$ and $P(t+dt)$ by
similar procedures to those of previous sections.
Since, in the above case, we have $\mbox{Tr} P(t)=N$ and
%
% 5-2
%
$$
\mbox{Tr}(P(t)P(t+dt))=\sum_{i=1}^N \kappa_i(t), \eqno(5.3)$$
(5.2) can be rewritten as
$$
d([\Psi(t)],[\Psi(t+dt)])=\sqrt{2\mbox{Tr}(P(t)\{P(t)-P(t+dt)\})}.
\eqno(5.4)$$
Eqs. (5.1),(5.4) and the relation $\mbox{Tr}([A,B])=0$ yield
$$
\eqalign {
d([\Psi(t)],[\Psi(t+dt)]) &=\frac{|dt|}{\hbar}
\sqrt{\mbox{Tr}(P(t)[H(t),[H(t),P(t)]])} \cr
&=\frac{|dt|}{\hbar} \mbox{Tr}([P(t), H(t)][H(t), P(t)]) \cr
&=\left\| \frac{dP(t)}{dt} \right\| |dt|. \cr
&=\| dP(t) \|. } \eqno(5.5)$$
It can be easily seen that the r.h.s.\ of (5.5) is proportional
to $\Delta {\cal E}_N(t)$ defined by (1.12).  Now we are led to
$$
d([\Psi(t)],[\Psi(t+dt)])=\frac{\sqrt2}{\hbar} \Delta {\cal E}_N(t)|dt|.
\eqno(5.6)$$
For finitely separated $[\Psi(t_1)]$ and $[\Psi(t_2)]$ in
$G_N$, the triangle inequality (2.21) implies
$$
\int_{t_1}^{t_2} \Delta {\cal E}_N(t)dt \geq \frac{\hbar}{\sqrt2}
d([\Psi(t_1)],[\Psi(t_2)]), t_2 \geq t_1 . \eqno(5.7)$$
The formula (3.19) then leads us to (1.13) or (1.14) or (1.15).
%
% 6-1
%
\newpage
\noindent
{\bf $\S$6. Discussions}
\vspace{0.5cm}

In this paper, we have been mainly engaged in obtaining the formula
(3.19) for the distance in $G_N$.  As an application of it,
we obtained the generalized version of the time--energy
uncertainty relation, (1.13), of Anandan--Aharonov type.
For $N$=1, it reduces to the result of Anandan and Aharonov, (1.7).

Our definition of the distance
$d([\Psi],[\Phi])$ in $G_N$, (2.16), is intimately related
to that in $\Omega$. The discussions around (4.13) or (5.5)
indicate that the distance between two infinitesimally
separated $N$-planes $[\Psi]$ and $[\Psi + d\Psi]$ is given by
$$
d([\Psi],[\Psi + d\Psi])= \| dP_\Psi \|, \eqno(6.1)$$
$$
dP_\Psi = P_{\Psi + d\Psi} - P_\Psi, \eqno(6.2)$$
where $P_{\Psi + d\Psi}$ and $P_\Psi$ are projection
operators associated with the $N$-frames $\Psi + d\Psi$
and $\Psi$, respectively. If we make use of some real
coordinates $s=(s^1, s^2, \ldots)$ to specify $[\Psi]$ or
$P_\Psi$, we have
$$
\{d([\Psi],[\Psi + d\Psi])\}^2 =
G_{\mu\nu}(s)ds^\mu ds^\nu, \eqno(6.3) $$
with $G_{\mu\nu}(s)$ given by (4.14).
Metric tensors of this kind have been discussed recently
by two of the present authors.$^{7)}$
We stress that the metric tensor and the geodesic for the
Grassmannian $G_N$ can be simply expressed in terms of
projection operators belonging to $\wedge_N$.

When the Hamiltonian $H$ is independent of time, the
$\Delta {\cal E}_N(t)$ in (1.12) does not depend on $t$ and is
determined only through $[\Psi(0)]$.  We suppose that
$\Psi(t)$ is orthogonal to $\Psi(0), i.e.,  \< \psi_i(t)|
\psi_j(0) \> =0, i,j=1,2,\ldots,N$ and that $\Delta {\cal E}_N(0)$ is
nonvanishing.  Then we have $\kappa_i=0, i=1,2,\ldots,N$,
and hence
$$
t\Delta {\cal E}_N(0)\geq \pi \sqrt N \frac{\hbar}2. \eqno(6.5)$$
In other words, an $N$-frame $\Psi$ needs $\pi\sqrt N \hbar/
(2 \Delta {\cal E}_N(0))$ of time to develop to the one orthogonal
to the original one.

Other applications of the distance formula (3.19)
will be discussed elsewhere.
%
% 6-2
%
\newpage
\noindent
\underline{\bf Acknowledgements}
\vspace{0.3cm}

The authors are grateful to Shinji Hamamoto, Shinobu Hosono and
Hitoshi Yamakoshi for valuable discussions.
One of the author (M.H.) thanks Yoshiyuki Watanabe and Kazuyuki Fujii
for calling his attention to Refs.3) and 6), respectively.
This work is partially supported by the Grant--in--Aid
\# 04804014 of the Ministry of Education, Science and Culture.
\vspace{1cm}\\

%
%   A-1
%
\newpage
\noindent
{\bf Appendix A.\ Geodesic and one parameter subgroup in $\Omega$}
\vspace{0.5cm}
\\
(a) In the following,we discuss that a geodesic in $\Omega$,
$$
\Gamma = \{ \gamma(t):t \in \R,\gamma(0) = \1(\mbox{unit operator})\},
\eqno(A.1) $$
can be identified with a subgroup
$ \{e^{itY}:Y=Y^\dagger,t \in \R \}$ of $\Omega$.
More systematic expositions can be found in Ref. 5.
\\
(b)
For any $\omega \in \Omega$,we define the mapping $I_\omega$ as follows.
For arbitrary $\omega, \omega^\prime \in \Omega$,
we can think of a geodesic $\Gamma$ such that
$\omega=\gamma(a)$ and $\omega^\prime =\gamma(b),
a,b \in \R.$
We define $I_\omega$ by
$$
I_\omega(\gamma(t)) = \gamma(2a-t).
\eqno(A.2)$$
Eqs. (2.10) and (2.12) imply that
$\|\frac{d\gamma(t)}{dt}\|$ is constant on $\Gamma$.
Then,we can characterize the mapping $I_\omega$ by
$$ I_\omega(\omega)=\omega, \eqno(A.3)$$
$$
\left\{\frac{d}{dt}I_\omega(\gamma(t))\right\}_{t=a}
= -\left( \frac{d\gamma(t)}{dt}\right)_{t=a}
\eqno(A.4)
$$
and
$$
\|\frac{d}{dt} I_\omega(\gamma(t))\| = \left\|\frac{d\gamma(t)}{dt}
\right\|,t \in \R.
\eqno(A.5)$$
%
% A-2
%
It is clear that $I_\omega(\omega^\prime)$ is well-defined for any
$\omega^\prime \in \Omega$ and that $I_\omega$ is unique.
\\
(c) On the other hand,if we define the mapping $J_\omega$ by
$$
J_\omega(\omega^\prime) =\omega(\omega^\prime)^{-1}\omega,
\ \ \omega,\omega^\prime \in \Omega,
\eqno(A.6)$$
we have
$$ J_\omega(\omega)=\omega, \eqno(A.7)$$
$$
\eqalign{
\left\{\frac{d}{dt}J_\omega(\gamma(t))\right\}_{t=a}
&= \left\{\frac{d}{dt}(\omega(\gamma(t))^{-1}\omega)\right\}_{t=a}
\cr
&=-\{\omega\gamma(t)^{-1}\frac{d\gamma(t)}{dt}\gamma(t)^{-1}\omega\}_{t=a}\cr
&= -(\frac{d\gamma(t)}{dt})_{t=a}}
\eqno(A.8)$$
and
$$
\left\| \frac{d}{dt}J_\omega(\gamma(t))\right\|
= \left\|-\omega\gamma(t)^{-1}\frac{d\gamma(t)}{dt}
\gamma(t)^{-1}\omega \right\|
= \left\|\frac{d\gamma(t)}{dt}\right\|, t \in \R.
\eqno(A.9)$$
Thus,we can identify $J_\omega$ with $I_\omega$:
$$
J_\omega = I_\omega.
\eqno(A.10) $$
\\
(d)Now,we assume that the unit operator \1 lies on $\Gamma$
 and choose the parameter t such that
$\gamma(0) = \1$.
We have
%
% A-3
%
$$
I_{\gamma(t)}(I_{\gamma(0)}(\gamma(u)))
= I_{\gamma(t)}(\gamma(-u)) = \gamma(2t+u)
\eqno(A.11)$$
and
$$
J_{\gamma(t)}(J_{\gamma(0)}(\gamma(u)))= J_{\gamma(t)}(\gamma(u)^{-1})
=\gamma(t)\gamma(u)\gamma(t).
\eqno(A.12)$$
{}From (A.10-12), we obtain
$$
\gamma(2t+u) =\gamma(t)\gamma(u)\gamma(t)
\eqno(A.13)$$
Putting $u=0,t,2t,3t,\ldots,$ in (A.13),
we have
$$
\gamma(nt) = \{\gamma(t)\}^n,n=0,1,2,\ldots.
\eqno(A.14)$$
Since $\gamma(t)$ should be continuous in t,
we conclude that $\gamma(t)$ can be written as
$$
\gamma(t)=e^{itY}, \quad t \in \R, \quad Y=Y^\dagger. \eqno(A.15)$$
We thus arrive at the statement in (a).
%
%     B-1
%
\vspace{2.0cm}\\
{\bf Appendix B. \ Calculation of $\mbox{Tr}(e^{iY}P(1)e^{-iY}P(0))$}
\vspace{0.5cm}

In this Appendix,we calculate $\mbox{Tr}(e^{iY}P(1)e^{-iY}P(0))$
through several steps,
where $Y$ is the operator given by (3.2).
We use the abbreviated notations
$P_0=P(0),P_1=P(1),x=P(0)P(1),y=P(1)P(0)$.
%
%   B-2
%
\\
(a) Making use of the relations
$$
\eqalign{
&xP_1=x,P_0x=x,\cr
&y^nP_1=P_1x^n,y^nx=P_1x^n,n=1,2,3,\ldots,
}\eqno(B.1)$$
repeatedly,we find that $Y$ in (3.2) satisfies
$$
Y^nP_1 = f_n(x) + P_1g_n(x),n=0,1,2,\ldots,
\eqno(B.2)$$
with $f_n(x)$ and $g_n(x)$ given by
$$
\left(
\begin{array}{c}
f_n(x) \\ g_n(x)
\end{array}\right)
 =\{Q(x)\}^n
\left(
\begin{array}{c}
0 \\ 1\end{array}\right),
n=0,1,2,\ldots,
\eqno(B.3) $$
$$
Q(x) = \left(
\begin{array}{cc}
\alpha(x)+i\beta(x)x   &\{\alpha(x)+i\beta(x)\}x \\
-(\alpha(x)+i\beta(x)) &-(\alpha(x)+i\beta(x)x)
\end{array}\right).
\eqno(B.4)$$
Hence,we have
$$
e^{iY}P_1= (1\ \ P_1)e^{iQ(x)}\left(
\begin{array}{c}
0 \\ 1 \end{array}\right).
\eqno(B.5)$$
Similarly,we obtain
$$
Y^nP_0 = {\tilde f}_n(y) + P_0{\tilde g}_n(y),
\eqno(B.6)$$
$$
\left(\begin{array}{c}
{\tilde f}_n(y) \\ {\tilde g}_n(y) \end{array}\right)
 = \{{\tilde Q}(y)\}^n
\left(\begin{array}{c}
0 \\ 1 \end{array}\right),
n=0,1,2,\ldots,\eqno(B.7)$$
$$
{\tilde Q}(y) = \left(\begin{array}{cc}
-(\alpha(y)+i\beta(y)y) & -(\alpha(y) + i\beta(y))y \\
\alpha(y) + i\beta(y)  & \alpha(y) + i\beta(y)y
\end{array}\right),\eqno(B.8)$$
and
$$
e^{-iY}P_0 = (1\ \ P_0)e^{-i{\tilde Q}(y)}\left(\begin{array}{c}
0 \\ 1\end{array}\right).\eqno(B.9)$$
%
%    B-3
%
\\
(b) From (B.2) and (B.6),we obtain
$$
\mbox{Tr}(e^{iY}P_1e^{-iY}P_0) = \sum_{m,n=0}^{\infty}\frac{i^m(-i)^n}
{m!n!}G_{mn},\eqno(B.10)$$
where $G_{mn}$ is given by
$$
G_{mn} = \mbox{Tr}(\{f_m(x)+P_1g_m(x)\}
\{{\tilde f}_n(y) + P_0{\tilde g}_n(y)\}).
\eqno(B.11)$$
Noticing that (B.3) and (B.7) ensure the existence of $h_m(x)$
and ${\tilde h}_n(y)$ such that
$$
f_m(x) = xh_m(x) , {\tilde f}_n(y) = y{\tilde h}_n(y),n=0,1,2,\ldots,
\eqno(B.12)$$
we obtain
$$
G_{mn} = \mbox{Tr}(y\{h_m(y)+g_m(y)\}\{{\tilde h}_n(y) +
{\tilde g}_n(y)\}).
\eqno(B.13)$$
In deriving (B.13),the following relations have been made use of:
$$
\eqalign{
\mbox{Tr}(x^{p+1}y^{q+1})
&=\mbox{Tr}(P_1x^py^{q+1})
=\mbox{Tr}(x^{p+1}P_0y^q)\cr
&=\mbox{Tr}(P_1x^pP_0y^q)
=\mbox{Tr}(y^{p+q+1})
=\mbox{Tr}(x^{p+q+1}),\cr
&p,q=0,1,2,\ldots,
}\eqno(B.14)$$
\\
(c) From the definitions of $P_0$,$P_1$ and $y$,we have
%
%   B-4
%
$$
y= \sum_{i,j=1}^N |\psi_i(1) \> a_{ij} \< \psi_j(0)|,
\eqno(B.15)$$
$$a_{ij} =  \< \psi_i(1)|\psi_j(0) \> .
\eqno(B.16)$$
We then have
$$
y^{q+1} = \sum_{i,j=1}^N |\psi_i(1) \> (AK^q)_{ij} \<
\psi_j(0)|,q=0,1,2,\ldots,
\eqno(B.17)$$
where $A$ is the $N \times N$ matrix whose $ij$-element is $a_{ij}$
and the $N \times N$ matrix $K$
is defined by
$$
K = A^\dagger A.
\eqno(B.18)$$
{}From (B.17),we obtain
$$
\mbox{Tr}(y^q) = \mbox{tr} (K^q),q=1,2,3,\ldots,
\eqno(B.19)$$
where $\mbox{tr}$ denotes the trace of an $N \times N$ numerical matrix.
Eqs. (B.10),(B.13) and (B.19) yield
$$
\mbox{Tr}(e^{iY}P_1e^{-iY}P_0) = \mbox{tr}(KG(K){\tilde G}(K)),
\eqno(B.20)$$
where $G(K)$ and ${\tilde G}(K)$ are given by
$$
G(K) = \sum_{n=0}^{\infty}\frac{i^n}{n!}
(h_n(K) + g_n(K)) = (K^{-1}\ \ 1)e^{iQ(K)}\left(
\begin{array}{c}
0 \\ 1 \end{array}\right),
\eqno(B.21a)$$
$$
{\tilde G}(K) = \sum_{n=0}^{\infty}\frac{(-i)^n}{n!}
({\tilde h}_n(K) + {\tilde g}_n(K)) = (K^{-1}\ \ 1)e^{-i{\tilde Q}(K)}\left(
\begin{array}{c}
0 \\ 1 \end{array}\right),
\eqno(B.21b)$$
The apperrance of $K^{-1}$ in (B.21) is only spurious and no
%
%    B-5
%
dangerous procedure is involved there.
\\
(d) If we denote the $N$-th unit matrix by $1_N$,$Q(K)$ can be written as
$$
Q(K) = \sum_{i=1}^3 q_i(K)\sigma_i,
\eqno(B.22)$$
$$
\eqalign{
&q_1(K) = \frac12(K-1_N)(\alpha(K)+i\beta(K)),\cr
&q_2(K) = \frac i2(K+1_N)(\alpha(K)+i\beta(K)),\cr
&q_3(K) = \alpha(K)+iK\beta(K).
}\eqno(B.23)$$
We now have
$$
G(K) = \cos\sqrt{s(K)} + (K-1_N)\beta(K)\frac{\sin\sqrt{s(K)}}{\sqrt{s(K)}}
\eqno(B.24)$$
with $s(K)$ given by
$$
s(K) = \sum_{i=1}^3 \{q_i(K)\}^2.
\eqno(B.25)$$
Similar manipulations yield
$$
{\tilde G}(K)=G(K).
\eqno(B.26)$$
{}From (B.24),(B.26) and (B.20),we obtain
$$
\mbox{Tr}(e^{iY}P_1e^{-iY}P_0) =
\mbox{tr}\left(K\left\{\cos\sqrt{s(K)} + (K-1_N)\beta(K)
\frac{\sin\sqrt{s(K)}}{\sqrt{s(K)}}\right\}^2\right),
\eqno(B.27)$$
yielding (3.3).

%
%     C-1
%
\vspace{2.0cm}
\noindent
{\bf Appendix C. \ Proof of (3.6)}
\vspace{0.5cm}

It is clear that $K$ is hermitian and positive definite so that we
have $\kappa_i \geq 0,i=1,2,\ldots,N$.
Supposing that $uKu^\dagger$,$u \in U(N)$,is diagonal,we obtain
$$
\sum_{k,l=1}^N u_{ik}(\delta_{kl}-K_{kl})(u^\dagger)_{lj}
 = (1-\kappa_i)\delta_{ij} =  \< \varsigma_i|\varsigma_j \> ,
\eqno(C.1)$$
where $|\varsigma_i \> $ is given by
$$
|\varsigma_i \>  = \sum_{l=1}^N (u_{il})^*
|\eta_l \> ,
\eqno(C.2)$$
$$
|\eta_i \>  = |\psi_i(0) \> -\sum_{k=1}^N  \< \psi_k(1)|\psi_i(0) \>
|\psi_k(1) \> .
\eqno(C.3)$$
Putting $i=j$ in (C.1),we have $\kappa_i \leq 1,i=1,2,\ldots,N$.
%
%     D-1
%
\newpage
\vspace{2.0cm}
\noindent
{\bf Appendix D. \ $e^{iG(P_0, P_1)}$ and $U$ }
\vspace{0.5cm}

{}From (3.18), we have
$$
e^{iG(P_0, P_1)}=\sum_{n=0}^\infty \frac 1{n!} J^n L^n, \eqno(D.1)$$
$$
J=\sum_{n=0}^\infty \frac{(2n)!!}{(2n+1)!!}(P_0-P_1)^{2n}, \eqno(D.2)$$
$$
L=P_0P_1-P_1P_0=-(P_0-P_1)(1-P_0-P_1) \eqno(D.3)$$
since $J$ commutes with $L$. By making use of the relations$^{6)}$
$$
\{ P_0-P_1, 1-P_0-P_1 \}=0, \eqno(D.4)$$
$$
(P_0-P_1)^2+(1-P_0-P_1)^2=1 \eqno(D.5)$$
and
$$
(P_0-P_1)J=\frac{\mbox{Arcsin}(P_0-P_1)}{\sqrt{1-(P_0-P_1)^2}}, \eqno(D.6)$$
we are led to
$$
\eqalign {
e^{iG(P_0, P_1)}&=\cos(\mbox{Arcsin}(P_0-P_1))-\sin(\mbox{Arcsin}(P_0-P_1))
\frac{1-P_0-P_1}{\sqrt{1-(P_0-P_1)^2}} \cr
&=\sqrt{1-(P_0-P_1)^2}-\frac{(P_0-P_1)(1-P_0-P_1)}
 {\sqrt{1-(P_0-P_1)^2}}. } \eqno(D.7)$$
It is easy to derive (3.21) from (D.7).
%
%     E-1
%
\newpage
\vspace{2.0cm}
\noindent
{\bf Appendix E. \ Proof of (4.1)}
\vspace{0.5cm}

The definitions (4.2) of $\sigma$ and $\lambda$ mean that
$|\psi \> $ and $|\phi \> $ can be expressed as
$$
|\psi \>  = \sqrt{\sigma}e^{i\delta}|\xi \>  + |\varsigma \>
\eqno(E.1)$$
$$
|\phi \>  = \sqrt{\lambda}e^{i\epsilon}|\xi \>  + |\eta \>
\eqno(E.2)$$
where $\delta,\epsilon \in \R$
and $|\varsigma \> $ and $|\eta \> $ are appropriate vectors orthogonal to
$|\xi \> $.
Since $|\psi \> $ and $|\phi \> $ are unit vectors,we have
$$
 \< \varsigma|\varsigma \>  = 1-\sigma, \< \eta|\eta \>  = 1-\lambda
\eqno(E.3)$$
and
$$
| \< \eta|\varsigma \> | \leq \sqrt{1-\sigma}\sqrt{1-\lambda}.
\eqno(E.4)$$
The lower bound of $\sqrt{\kappa}$ is estimated as
$$
\eqalign{
\sqrt{\kappa} &= | \< \phi|\psi \> |\cr
&=|(\sqrt{\lambda}e^{-i\epsilon} \< \xi| +  \<
\eta|)(\sqrt{\sigma}e^{i\delta}|\xi \> + |\varsigma \> )|\cr
&=|\sqrt{\lambda\sigma}e^{i\delta-i\epsilon} +  \< \eta|\varsigma \> |\cr
&\geq \sqrt{\lambda\sigma}-| \< \eta|\varsigma \> |\cr
&\geq \sqrt{\lambda\sigma}-\sqrt{1-\sigma}\sqrt{1-\lambda}
}\eqno(E.5)$$
Noticing the formula
$\mbox{Arccos}(\sqrt{\lambda\sigma}-\sqrt{1-\lambda}\sqrt{1-\sigma})
 = \mbox{Arccos} \sqrt{\sigma} + \mbox{Arccos}\sqrt{\lambda}$,
we obtain (4.1).
%
%    F-1
%
\newpage
\vspace{2.0cm}
\noindent
{\bf Appendix F. \ Proof of (4.20)}
\vspace{0.5cm}

We consider a two-parameter family of unitary operators
$$
W(r,t)=e^{irY(t)}=(s^1(r,t,), s^2(r,t,),\ldots), \eqno(F.1)$$
where $s^\mu(r,t), \mu=1,2,\ldots$, are local coordinates of
$\Omega$ and $Y(t)$ is so chosen that
$$
\frac d{dt}\|Y(t)\|=0. \eqno(F.2)$$
We note that (F.2) is equivalent to
$$
\frac \partial{\partial t}
\left \| \frac{\partial W(r,t)}{\partial r} \right \|^2=0. \eqno(F.3)$$
Given $g_{\mu\nu}(s)$ by (2.8), we define $A(r,t)$ by
$$
A(r,t)=g_{\mu\nu}(s)\frac{\partial s^\mu(r,t)}{\partial r}
\frac{\partial s^\nu(r,t)}{\partial t} . \eqno(F.4)$$
We have
$$
A(0,t)=0 \eqno(F.5)$$
since $W(0,t)=\1$ and $s^\mu(0,t)$ is independent of $t$.
On the other hand, with the help of the formula
$\partial_\kappa g_{\mu\nu}(s)=\Gamma_{\kappa\mu}^\rho(s)
g_{\rho\nu}(s)+\Gamma_{\kappa\nu}^\rho(s)g_{\rho\mu}(s)$,
we obtain
%
%    F-2
%
$$
\frac\partial{\partial r}A(r,t)=g_{\mu\nu}(s)
\left\{\frac{\partial s^\mu(r,t)}{\partial t}K^\nu(r,t)+
\frac{\partial s^\mu(r,t)}{\partial r}L^\nu(r,t) \right\}, \eqno(F.6)$$
where $K^\nu(r,t)$ and $L^\nu(r,t)$ are given by
$$
K^\nu(r,t)=\frac{\partial^2 s^\nu(r,t)}{\partial r^2}+
\Gamma_{\kappa \lambda}^\nu(s) \frac{\partial s^\kappa(r,t)}{\partial r}
\frac{\partial s^\lambda(r,t)}{\partial r}, \eqno(F.7)$$
$$
L^\nu(r,t)=\frac{\partial^2 s^\nu(r,t)}{\partial r \partial t}+
\Gamma_{\kappa \lambda}^\nu(s) \frac{\partial s^\kappa(r,t)}{\partial r}
\frac{\partial s^\lambda(r,t)}{\partial t}. \eqno(F.8)$$
The fact that $\{ W(r,t): r \in \R, t;fixed \}$ is a geodesic in
$\Omega$ implies $K^\nu(r,t)=0$, while the assumption $(F.2)$ or
$(F.3)$ means that
$$
\frac{\partial}{\partial t}\left\{ g_{\mu\nu}(s)
\frac{\partial s^\mu(r,t)}{\partial r}\frac{\partial s^\nu(r,t)}{\partial r}
\right\}= g_{\mu\nu}(s) \frac{\partial s^\mu(r,t)}{\partial r}
L^\nu(r,t)=0. \eqno(F.9)$$
We now have
$$
\frac{\partial}{\partial r} A(r,t)=0. \eqno(F.10)$$
Eqs.\ $(F.5)$ and $(F.10)$ lead us to
$$
A(r,t)=0. \eqno(F.11)$$
For the $\gamma(t)$ in (4.5), we obtain
$$
\frac{d \gamma(t)}{dt}=\left(\frac{\partial W(r,t)}{\partial r}
\right)_{r=R(t)} \frac{dR(t)}{dt}+
\left(\frac{\partial W(r,t)}{\partial t}
\right)_{r=R(t)}. \eqno(F.12)$$
%
%     F-3
%
{}From $(F.11)$ and $(F.12)$, we see that
$$
\left\| \frac{d\gamma(t)}{dt} \right\|^2 =
\left| \frac{dR(t)}{dt} \right|^2
\left\| \frac{\partial W(r,t)}{\partial r} \right\|_{r=R(t)}^2 +
\left\| \frac{\partial W(r,t)}{\partial t} \right\|_{r=R(t)}^2. \eqno(F.13)$$
We conclude (4.20) since we know
$\left\| \frac{\partial W(r,t)}{\partial r} \right\| =\|Y(t)\|=\|Y(1)\| $.
We note that a more general version of the above discussion
can be found in Ref. 5.
\end{document}